\documentclass[aps,prd,preprint,superscriptaddress,nofootinbib,showpacs]{revtex4}
\usepackage{amsfonts}
\usepackage{mathrsfs}
\usepackage{amssymb}
\usepackage{amsmath}
\usepackage{graphicx,epsfig}

\def\bwt{\begin{widetext}}
\def\ewt{\end{widetext}}
\def\be{\begin{equation}}
\def\ee{\end{equation}}
\def\bea{\begin{eqnarray}}
\def\eea{\end{eqnarray}}
\def\bean{\begin{eqnarray*}}
\def\eean{\end{eqnarray*}}
\def\bary{\begin{array}}
\def\eary{\end{array}}
\def\bit{\begin{itemize}}
\def\eit{\end{itemize}}
\def\GeV{\rm GeV}

\def\su5u1{SU(5) \times U(1)}
\def\fsu5u1{SU(5) \times U(1)'}
\def\so10{SO(10)}
\def\sq20{SO(10) \times SO(10)}

\usepackage{amssymb}

\begin{document}

\setlength{\parskip}{0cm}

\title{Toward the Natural and Realistic NMSSM with and without $R$-Parity}

\author{Taoli Cheng}

\affiliation{State Key Laboratory of Theoretical Physics
and Kavli Institute for Theoretical Physics China (KITPC),
Institute of Theoretical Physics, Chinese Academy of Sciences,
Beijing 100190, P. R. China}

\author{Jinmian Li}

\affiliation{State Key Laboratory of Theoretical Physics
and Kavli Institute for Theoretical Physics China (KITPC),
Institute of Theoretical Physics, Chinese Academy of Sciences,
Beijing 100190, P. R. China}

\author{Tianjun Li}

\affiliation{State Key Laboratory of Theoretical Physics
and Kavli Institute for Theoretical Physics China (KITPC),
Institute of Theoretical Physics, Chinese Academy of Sciences,
Beijing 100190, P. R. China}

\affiliation{George P. and Cynthia W. Mitchell Institute for
Fundamental Physics and Astronomy, Texas A$\&$M University,
College Station, TX 77843, USA}

\author{Xia Wan}

\affiliation{Institute of Theoretical Physics $\&$ State Key Laboratory of
Nuclear Physics and Technology, Peking University, Beijing 100871,
P. R. China}

\author{You kai Wang}

\affiliation{State Key Laboratory of Theoretical Physics
and Kavli Institute for Theoretical Physics China (KITPC),
Institute of Theoretical Physics, Chinese Academy of Sciences,
Beijing 100190, P. R. China}

\author{Shou-hua Zhu}

\affiliation{Institute of Theoretical Physics $\&$ State Key Laboratory of
Nuclear Physics and Technology, Peking University, Beijing 100871,
P. R. China}

\affiliation{Center for High Energy Physics, Peking University,
Beijing 100871, P. R. China }



\begin{abstract}

From the current ATLAS and CMS results on Higgs boson mass 
and decay rates, the NMSSM is obviously better than the MSSM.
To explain the fine-tuning problems such as gauge hiearchy 
problem and strong CP problem in the SM, we point out that
supersymmetry does not need to provide a dark matter candidate, 
{\it i.e.}, $R$-parity can be violated. Thus, we consider 
three kinds of the NMSSM scenarios: in Scenarios I and II  
$R$-parity is conserved and the lightest neutralino relic 
density is respectively around and smaller than the observed 
value, while in Scenario III $R$-parity is violated. To fit 
all the experimental data, we consider the $\chi^2$ analyses, 
and find that the Higgs boson mass and decay rates can be
explained very well in these Scenarios. Considering the 
small $\chi^2$ values and fine-tuning around 2-3.7\% (or 1-2\%), 
we obtain the viable parameter space with light (or relatively 
heavy) supersymmetric particle spectra only in Scenario III
(or in Scenarios I and II). Because the singlino, Higgsinos, 
and light stop are relatively light in general, we can relax
the LHC supersymmetry search constraints but the XENON100 
experiment gives a strong constraint in Scenarios I and II.
In all the viable parameter space, the anomalous magnetic 
moment of the muon $(g_{\mu} - 2)/2$ are generically small.
With $R$-parity violation, we can increase $(g_{\mu} - 2)/2$,
and avoid the contraints from the LHC supersymmetry searches 
and XENON100 experiment. Therefore, Scenario III with  
$R$-parity violation is more natural and realistic 
than Scenarios I and II.

\end{abstract}



\maketitle

\section{Introduction}

The Higgs boson mass in the Standard Model (SM) is not
stable against qunatum corrections and its square has quadratic divergences.
Because the reduced Planck scale is about 16 order larger than the electroweak
(EW) scale, there exists huge fine-tuning around $10^{-32}$ to obtain the EW-scale
Higgs boson mass. Supersymmetry is a symmetry
between the bosonic and fermionic states, and it naturally solves this problem due to
the cancellations between the bosonic and fermionic quantum corrections.
 In Minimal Supersymmetric SM (MSSM),
 the  gauge couplings for $SU(3)_C$, $SU(2)_L$ and $U(1)_Y$
gauge symmetries are unified at about $2\times 10^{16}$~GeV~\cite{Ellis:1990zq},
which strongly suggests Grand Unified Theories (GUTs).
Unlike the SM, we can have the renormalizable superpotential terms that
violate the baryon and lepton numbers, and then there may exist proton decay
problem. To solve such problem, we usually introduce the $R$-parity under which
the SM particles are even while the extra supersymmetric particles (sparticles)
are odd. Thus,
the lightest supersymmetric particle (LSP) like neutralino can
be cold dark matter candidate~\cite{Ellis:1983wd, Goldberg:1983nd}.

However, there are strong constrains on the supersymmetry viable parameter
space from the recent LHC supersymmetry
searches~\cite{Aad:2011ib, Aad:2011qa, Chatrchyan:2011zy, SUSY-ATLAS-CMS, LHC-SUSY-EX}.
For example, in the Minimal Supergravity (mSUGRA) model or Constrained MSSM (CMSSM),
gluino mass should be larger
than about 1.4 TeV and 850 GeV for squark masses around and
much larger than gluino mass, respectively. Also,
 squarks (at least the first two generation squarks) must have masses
larger than about 1.1 TeV from the ATLAS and CMS Collaborations
at the LHC~\cite{Aad:2011ib, Aad:2011qa, Chatrchyan:2011zy, SUSY-ATLAS-CMS, LHC-SUSY-EX}.

Recently, the ATLAS  and CMS Collaborations have announced the discovery
of a Higgs-like boson with mass around $126.5$~GeV and $125.3 \pm 0.6$~GeV,
respectively~\cite{HiggsDiscovery,atlashiggs,cmshiggs}.
In the $\gamma \gamma $ final state, the ATLAS and CMS rates are
roughly $1.9\pm 0.5$ and $1.56\pm 0.43$  times the SM prediction.
In the $ZZ\to 4\ell$ channel, the ATLAS and CMS signals are roughly
$1.1^{+0.5}_{-0.4}$ and $0.7^{+0.4}_{-0.3}$ times the SM prediction, respectively.
In the $b {\bar b}$, $\tau^+\tau^-$ and $WW\to \ell\nu\ell\nu$ channels,
the ATLAS rates are respectively $0.48^{+2.17}_{-2.12}$, $0.16^{+1.72}_{-1.84}$,
and $0.52^{+0.57}_{-0.60}$ times the SM prediction, and
the CMS rates are respectively $0.15^{+0.73}_{-0.66}$, $-0.14^{+0.76}_{-0.73}$,
and $0.62^{+0.43}_{-0.45}$. So these rates are somewhat suppressed compare to
the SM prediction but error bars are relatively large.
The Higgs physics implications in the supersymmetric SMs (SSMs)
have been studied extensively~\cite{SUSY-PH, Carena:2011aa, Ellwanger:2012ke,
N-SUSY-PH, Gunion:2012gc}.
By the way, the new results from the CDF and D0 experiments~\cite{newtevatron}
support the $\sim 125~{\rm GeV}$ Higgs signal and suggest an enhancement relative to
the SM of the $W$+Higgs with Higgs$\to b{\bar b}$ rate by a factor of $1.97^{+0.74}_{-0.68}$.
But we will consider not it here since it is different from the ATLAS and CMS results.

As we know, there are two Higgs doublets $H_u$ and $H_d$ in the MSSM
that gives masses  to the up-type quarks and down-type quarks/charged leptons,
respectively. The lightest CP-even
Higgs boson mass, which is a linear combination of $H_u^0$ and $H_d^0$ and
 usually SM-like, is smaller than $Z$ boson mass $M_Z$
at tree level. Thus, to realize
 the lightest CP-even Higgs boson mass around 125.5 GeV radiatively, the squark and/or gluino
masses will be about a few TeV in general in the mSUGRA/CMSSM. And then
there exists at least less than one-percent fine-tuning.
Moreover,  it is difficult to explain
the above Higgs decay rates and generate the correct Higgs boson 
mass simultaneously in the MSSM.
For example, if the SM-like Higgs particle has dominant component
from $H_u^0$, we can suppress the rates in the
$b {\bar b}$ and $\tau^+\tau^-$ final states, and then increase the
 $\gamma \gamma $ rate. But the rates for the $ZZ\to 4\ell$ and
$WW\to \ell\nu\ell\nu$ channels will increase as well.
Also, if the stop is light, we can increase the Higgs to two photon rate,
but it is difficult to generate the 125.5~GeV Higgs boson
mass~\cite{SUSY-PH, Carena:2011aa, Ellwanger:2012ke,
N-SUSY-PH, Gunion:2012gc}.
The possible model might be the light stau scenario~\cite{Carena:2011aa}.
Therefore, we shall consider the next to the MSSM (NMSSM) where
an SM singlet field $S$ is introduced. The points are the following:
(1) We can increase the Higgs quartic coupling from the superpotential
term $\lambda S H_d H_u$ if the ratio
$\tan\beta \equiv \langle H_u^0 \rangle/\langle H_d^0 \rangle$ of
the vacuum expectation values (VEVs)
for $H_u^0$ and $H_d^0$  is not large; (2)
We can suppress the couplings between the $W/Z$ gauge bosons and the Higgs
particle due to the mixings among $S$, $H_u^0$, and $H_d^0$.

On the other hand, the strong CP problem is another
big fine-tuning problem in the SM. From the experimental
bound on the neutron electric dipole moment (EDM), the strong CP phase
$\overline{\theta}$ is required to be smaller than $10^{-10}$.
An elegant and popular solution
to the strong CP problem is provided by the Peccei--Quinn (PQ)
mechanism~\cite{PQ}, in which a global axial symmetry $U(1)_{PQ}$ is
introduced and broken spontaneously at some high energy scale.
The axion $a$ is a pseudo-Goldstone boson
from the spontaneous $U(1)_{PQ}$ symmetry breaking, with a decay constant $f_a$.
The original Weinberg--Wilczek axion~\cite{WW} is excluded by experiment, in
particular by the non-observation of the rare decay $K \rightarrow \pi +
a$~\cite{review}. There are two viable ``invisible'' axion models in which the
experimental bounds can be evaded: the Kim--Shifman--Vainshtein--Zakharov
(KSVZ) axion model~\cite{KSVZ} and the
Dine--Fischler--Srednicki--Zhitnitskii (DFSZ) axion model~\cite{DFSZ}.
From laboratory, astrophysics, and cosmological constraints, the $U(1)_{PQ}$
symmetry breaking scale $f_a$ is constrained to  the range $10^{10}~{\rm GeV}
\leq f_a \leq 10^{12}~{\rm GeV}$~\cite{review}. Interestingly,
for such  $f_a$ range, the invisible axion can be a good cold dark matter
candidate with correct relic density~\cite{review}.

Because axion can be the correct dark matter candidate, supersymmetry
may only need to solve the gauge hiearchy problem and realize gauge coupling
unification. Therefore, we consider three kinds of the NMSSM scenarios:
in Scenario I,  $R$-parity is conserved and the LSP neutralino relic density is
around the observed value;
in Scenario II, $R$-parity is conserved and the LSP neutralino relic density
is smaller than the observed value; in Scenario III, $R$-parity is violated and
then the LSP neutralino is not stable. In particular, Scenario III is very
interesting since it can not only avoid the constraints from the 
LHC supersymmetry searches and XENON100 experiment~\cite{Aprile:2011hi}, 
but also may relax the other phenomenological constraints.
Moreover, the proton decay problem can be solved by requiring the baryon
or lepton number conservation~\cite{Barbier:2004ez},
or by requiring the minimal flavour violation~\cite{Nikolidakis:2007fc}.

In this paper, we shall study the natural and realistic NMSSM. We first
briefly review the naturalness condition in the SSMs and discuss the
NMSSM with and without $R$-parity. To satisfy
 the phenomenological constraints and fit the experimental
data, we consider the $\chi^2$ analyses for all three kinds of Scenarios,
and find that we can indeed explain the Higgs boson mass and decay rates
very well. Considering the small $\chi^2$ values and fine-tuning around 2-3.7\%,
we obtain the viable parameter space with light (e.g.~less than around 900~GeV) 
supersymmetric particle spectra only in Scenario I.
For the small  $\chi^2$ values and fine-tuning around 1-2\%,
we get the viable parameter space with relatively heavy 
(e.g.~less than about 1.2 TeV) supersymmetric particle spectra.
In particular, the best benchmark point has almost minimal $\chi^2$ and 
 3.7\% fine-tuning in Scenario III.
The generic features for the viable parameter space with smaller $\chi^2$ are that
the light stop is around 500 GeV or smaller,  the singlino and Higgsino are light
chargino and neutralinos, the Wino-like chargino is heavy, and
the Bino-like and Wino-like neutralinos are
the second heaviest neutralino and heaviest neutralinos, respectively.
Thus, we find that the LHC supersymmetry
search constraints can be relaxed due to quite a few jets and/or leptons in
the final states in Scenarios I and II, but the XENON100 experiment still gives
strong constraint on the dark matter direct detection cross sections. Moreover,
the correct dark matter relic density can be realized in Scenario I as well.
In particular, $\tan\beta$ is not large and the second lightest CP-even
 Higgs particle is SM-like~\cite{Ellwanger:2012ke, Gunion:2012gc},
 which is helpful to increase the SM-like Higgs boson mass.
However, the additional
contributions to the anomalous magnetic moment of the muon $(g_{\mu} - 2)/2$
 are smaller than three sigma low bound~\cite{Hagiwara:2011af} in general due to
relatively small $\tan\beta$.
As we know, with  $R$-parity violation, we can escape the constraints from
the LHC supersymmetry searches and XENON100 experiment, and the $R$-parity violation 
superpotential term(s) may increase the  muon $(g_{\mu} - 2)/2$ and explain 
the neutrino masses and mixings. Therefore,  Scenario III with  $R$-parity violation 
is more natural and realistic than Scenarios I and II.

This paper is organized as follows. We explain the naturalness criteria in the SSMs
in Section II. We present the NMSSM with and without $R$-parity in Section III,
and the experimental constraints/data and numerical analyses in Section IV. 
Section V is our conclusion.

\section{Naturalness Criteria in the SSMs}

For the GUTs with gravity mediated supersymmetry breaking, the usual
quantitative measure $\Delta_{\rm FT}$ for fine-tuning is the maximum of
the logarithmic derivative of $M_Z$ with respect to all the fundamental
parameters $a_i$ at the GUT scale~\cite{Ellis:1986yg}
\begin{eqnarray}
\Delta_{\rm FT} ~=~ {\rm Max}\{\Delta_i^{\rm GUT}\}~,~~~
\Delta_i^{\rm GUT}~=~\left|{{\partial{\rm ln}(M_Z)}\over
{\partial {\rm ln}(a_i^{\rm GUT})}}\right|~.~\,
\label{BG-FT}
\end{eqnarray}
In the following numerical calculations, we will use this definition
to calculate the fine-tuning.

However, the above fine-tuning definition is a little bit abstract. Thus,
we shall present the concrete bounds on the $\mu$ term, third-generation squark
masses and gluino mass in the following~\cite{Kitano:2005wc, Papucci:2011wy}.
The SM Higgs-like particle $h$ in the MSSM is a linear combintation
of $H_u^0$ and $H_d^0$. To simplify the discussion on naturalness,
we can reduce the Higgs potential to
\begin{eqnarray}
V&=& \overline{m}^2_h |h|^2 + {{\lambda_h}\over 4} |h|^4~,~
\end{eqnarray}
where $\overline{m}^2_h$ is negative. Minimizing the Higgs potential,
we get the physical SM-like Higgs boson mass $m_h$
\begin{eqnarray}
m_h^2 = - 2 \overline{m}^2_h ~.~
\end{eqnarray}
So the fine-tuning measure can also be defined as~\cite{Kitano:2005wc}
\begin{eqnarray}
\Delta_{\rm FT} \equiv {{2 \delta \overline{m}^2_h}\over {m_h^2}}~.~
\end{eqnarray}

For a moderately large $\tan\beta\equiv \langle H_u^0 \rangle/\langle H_d^0 \rangle$,
for instance, $\tan\beta \ge 2$, we have
\begin{eqnarray}
\overline{m}^2_h & \simeq & |\mu|^2 + m^2_{H_u}|_{\rm tree}+m^2_{H_u}|_{\rm rad} ~,~\,
\end{eqnarray}
where $\mu$ is the supersymmetric bilinear mass between $H_u$ and $H_d$, and
$m^2_{H_u}|_{\rm tree}$ and $m^2_{H_u}|_{\rm rad}$ are
the tree-level and radiative contributions to the soft supersymmetry-breaking
 mass squared for $H_u$.
Therefore, we obtain the following concrete bounds~\cite{Papucci:2011wy}:

\begin{itemize}

\item The upper bound on the $\mu$ term is
\begin{eqnarray}
\mu \lesssim 400~\GeV\left(\frac{m_{h}}{125.5~\GeV}\right)
\left(\frac{\Delta_{\rm FT} ^{-1}}{5\%}\right)^{-1/2} ~.~\,
\end{eqnarray}
Thus,  the $\mu$ term should be small than about 400 GeV for $5\%$ fine-tuning.
Consequncely, the charged and neutral Higgsinos will be light. In the NMSSM,
we just change the $\mu$ term to the effective $\mu$ term  
$\mu_{\rm eff} \equiv \lambda \langle S \rangle$.

\item The one-loop radiative corrections to $m_{H_u}^2$ in the leading
logarithmic approximation
from the stop sector are
\begin{eqnarray}
\delta m_{H_{u}}^{2}|_{\rm stop}=-\frac{3}{8\pi^{2}}y_{t}^{2}\left(m_{\widetilde{Q}_{3}}^{2}+
m_{\widetilde{U}_3^c}^{2}+|A_{t}|^{2}\right){\rm ln}\left(\frac{\Lambda}{\rm TeV}\right)~,
\label{Eq:Top}
\end{eqnarray}
where $y_t$ is top Yukawa coupling, $m_{\widetilde{Q}_{3}}^{2}$
and $m_{\widetilde{U}_3^c}^{2}$ are supersymmetry
breaking soft masses for the third generation quark doublet and
right-handed stop, $A_t$ is the top trilinear soft term, and $\Lambda$ is the
effective supersymmetry breaking mediation scale. Thus, one obtains
\begin{eqnarray}
\label{eq:ft-stop}
\sqrt{m_{{\tilde{t}_1}}^2+m_{\tilde{t}_2}^{2}}
\lesssim 1.2~{\rm TeV} \frac{\sin\beta}{(1+{\rm x}_{t}^{2})^{1/2}}
\left(\frac{{\rm ln}\left(\Lambda/{\rm TeV}\right)}{3}\right)^{-1/2}
\left(\frac{m_{h}}{125.5~{\rm GeV}}\right)\left(\frac{\Delta_{\rm FT}^{-1}}{5\%}\right)^{-1/2} ,
\end{eqnarray}
where ${\rm x}_{t}=A_{t}/\sqrt{m_{{\tilde{t}_1}}^2+m_{\tilde{t}_2}^{2}}$,
and $\tilde{t}_1$ and $\tilde{t}_2$
are two stop mass eigenstates.
Therefore, we obtain $\sqrt{m_{{\tilde{t}_1}}^2+m_{\tilde{t}_2}^{2}} \le 1.2~{\rm TeV}$.
Also, we can require that the lighter sbottom mass be smaller than
$m_{\tilde{t_2}}$, which is automatically satisfied via an simple mathematical proof.

\item The two-loop radiative corrections to $m_{H_u}^2$ in the leading
logarithmic approximation from gluino are
\begin{eqnarray}
\delta m_{H_{u}}^{2}|_{\rm gluino} = -\frac{2}{\pi^{2}}y_{t}^{2}\left(\frac{\alpha_{s}}{\pi}\right)
|M_{3}|^{2}{\rm ln}^{2}\left(\frac{\Lambda}{\rm TeV}\right)\, ,
\end{eqnarray}
where $\alpha_s$ is the strong coupling, and $M_3$ is the gluino mass. Here, the contributions
from the mixed $A_{t}M_{3}$ term , which are relevant for large A-term, are neglected.
Thus, the bound on gluino mass is
\begin{eqnarray}
M_{3} \lesssim 1.8~{\rm TeV} ~\sin\beta \left(\frac{{\rm ln}\left(\Lambda/{\rm TeV}\right)}{3}\right)^{-1}\left(\frac{m_{h}}{125.5~{\rm GeV}}\right)\left(\frac{\Delta_{\rm FT}^{-1}}{5\%}\right)^{-1/2}\, .
\end{eqnarray}
So the gluino mass is lighter than about 1.8 TeV.

\end{itemize}

Therefore, the natural MSSM  and NMSSM should have relatively smaller (effective) $\mu$ term,
 stop masses as well as gluino mass. In this paper, we shall
not only use Eq.~(\ref{BG-FT}) to calculate the numerical values of the fine-tuning,
but also consider the following natural supersymmetry conditions:
\begin{itemize}

\item The $\mu$ term or effective $\mu$ term is smaller than 300 GeV.

\item The squar root  $M_{\tilde t} \equiv \sqrt{m_{{\tilde{t}_1}}^2+m_{\tilde{t}_2}^{2}}$
of the sum of the two stop mass squares
 is smaller than
1.2~TeV. Consequencely, we can show that the light sbottom mass is smaller than
$m_{\tilde{t}_2}$.

\item The gluino mass is lighter than 1.5 TeV.

\end{itemize}
However, such kind of the natural MSSM and NMSSM might be excluded by the LHC supersymmetry
searches and XENON100 dark matter direct detection. Thus, the $R$-parity violation might be 
needed for the natural MSSM and
NMSSM, and then supersymmetry only needs to solve the fine-tuning problem and explain
the gauge coupling unification.


\section{The NMSSM with and without $R$-Parity}

Let us explain the convention first. We denote the quark doublets,
right-handed up-type quarks, right-handed down-type quarks, lepton doublets,
and right-handed leptons as $Q_i$, $U_i^c$, $D_i^c$, $L_i$, and
$E_i^c$, respectively. We denote the $SU(3)_C$, $SU(2)_L$, and
$U(1)_Y$ gauginos as ${\widetilde G}^a$, ${\widetilde W}^a$,
and ${\widetilde B}$, respectively.
To solve the $\mu$ problem in the MSSM,  we introduce a SM singlet field $S$
and consider the NMSSM with $Z_3$ symmetry
which forbids the $\mu$ term. The superpotential in the
NMSSM is
\begin{eqnarray}
W_\mathrm{NMSSM} &=& y_{ij}^u Q_i U_j^c H_u +  y_{ij}^d Q_i D_j^c H_d +
 y_{ij}^l L_i E_j^c H_d + \lambda S H_d H_u + \frac{1}{3} \kappa S^3 ~,~\,
\end{eqnarray}
where $y_{ij}^u$, $ y_{ij}^d$, $y_{ij}^l$, $\lambda$, and $\kappa$
are Yukawa couplings. The effective $\mu$ term is obtained after
$S$ obtains a VEV, {\it i.e.},
$\mu_{\rm eff} \equiv \lambda \langle S \rangle $.

The supersymmetry breaking soft terms  are
\begin{eqnarray}
-{\cal L} &=& \frac{1}{2} \bigg[
 M_1 \tilde{B}  \tilde{B}
\!+\!M_2 \sum_{a=1}^3 \tilde{W}^a \tilde{W}_a
\!+\!M_3 \sum_{a=1}^8 \tilde{G}^a \tilde{G}_a   + \mathrm{H.C.}
\bigg]
+ m_{H_u}^2 | H_u |^2 + m_{H_d}^2 | H_d |^2
\nonumber \\ &&
+ m_{S}^2 | S |^2
+m_{{\widetilde Q}_i}^2 |{\widetilde Q}_i|^2
 +m_{{\widetilde U}^c_i}^2 |{\widetilde U}^c_i|^2
 +m_{{\widetilde D}^c_i}^2 |{\widetilde D}^c_i|^2
 +m_{{\widetilde L}_i}^2 |{\widetilde L}_i|^2
 +m_{{\widetilde E}^c_i}^2 |{\widetilde E}^c_i|^2
\nonumber \\ &&
+\Bigl[  y_{ij}^u A_{ij}^u Q_i U_j^c H_u +   y_{ij}^d A_{ij}^d Q_i D_j^c H_d +
 y_{ij}^l A_{ij}^l L_i E_j^c H_d + \lambda A_\lambda S H_d H_u
\nonumber \\ &&
+ \frac{1}{3} \kappa A_\kappa S^3 + \mathrm{H.C.} \Bigl]~.~
\end{eqnarray}

Similar to the MSSM, the Higgs
sector of the NMSSM is described by the following six parameters
\begin{eqnarray}
\lambda\ , \ \kappa\ , \ A_{\lambda} \ , \ A_{\kappa}, \ \tan\beta\ ,\ \mu_\mathrm{eff}\; .
\end{eqnarray}
And the supersymmetry breaking soft mass terms for the Higgs bosons $m_{H_u}^2$,
$m_{H_d}^2$ and $m_{S}^2$ are determined implicitely by $M_Z$,
$\tan\beta$ and $\mu_\mathrm{eff}$ via the Higgs potential minimization.

In addition, from the theoretical point of view, we usually have the family universal
squark and slepton soft masses in the string model building. Therefore,
as in the mSUGRA/CMSSM, we consider the following universal supersymmetry
breaking soft terms
\begin{eqnarray}
M_1 = M_2 = M_3 \equiv M_{1/2}\; ,
\end{eqnarray}
\begin{eqnarray}
m_{{\widetilde Q}_i}^2 = m_{{\widetilde U}^c_i}^2 = m_{{\widetilde D}^c_i}^2
= m_{{\widetilde L}_i}^2 = m_{{\widetilde E}^c_i}^2 \equiv M_0^2\; ,
\end{eqnarray}
\begin{eqnarray}
A_{ij}^u = A_{ij}^d = A_{ij}^l \equiv A_0\; .
\end{eqnarray}

We consider the NUH-NMSSM in this paper: the Higgs soft mass terms
$m_{H_u}^2$, $m_{H_d}^2$ and $m_{S}^2$ are allowed to be different from
$M_0^2$ (and determined implicitely as mentioned above), and the trilinear
couplings $A_{\lambda}$, $A_{\kappa}$ and $A_0$ are not universal.
Therefore, the complete parameter space is characterized by
\begin{eqnarray}
\lambda\ , \ \kappa\ , \ \tan\beta\ ,\
\mu_\mathrm{eff}\ , \ A_{\lambda} \ , \ A_{\kappa} \ , \ A_0 \ , \ M_{1/2}\ ,
\ M_0\; ,
\label{9parameters}
\end{eqnarray}
where the last five parameters are taken at the GUT scale.

Next, we consider the $R$-parity violation. The most general renormalizable, gauge and $Z_3$ invariant,
and $R$-parity odd superpotential terms in the NMSSM are~\cite{Barbier:2004ez}
\begin{eqnarray}
W_{\rm RPV} &=& \lambda_i S  L_i H_u + \frac{1}{2} \lambda_{ijk} L_i L_j E_k^c +
\lambda'_{ijk} L_i Q_j D_k^c + \frac{1}{2} \lambda{''}_{ijk} U^c_i D^c_j D_k^c ~,~\,
\label{RPV-1}
\end{eqnarray}
where $\lambda_i$, $\lambda_{ijk}$, $\lambda'_{ijk}$,
and $\lambda{''}_{ijk}$ are Yukawa couplings.
In the above Eq.~(\ref{RPV-1}), the first three
terms conserve the baryon number while violate the lepton number, and the
last term conserves the lepton number while violates the baryon number. Thus,
to forbid the proton decay, we require either baryon number conservation or
lepton number conservation, {\it i.e.}, we turn on either the first three terms
or the last term in the above superpotential~\cite{Barbier:2004ez}.
The alternative ways are to consider
the minimal flavour violation~\cite{Nikolidakis:2007fc} or discrete $Z_N$
$R$-symmetry~\cite{Dreiner:2012ae}. In particular,
the $ \lambda_{ijk}$ and $\lambda'_{ijk}$ terms can contribute to the
anomalous magnetic moment of the muon $(g_{\mu} - 2)/2$ and generate the neutrino masse and mixings,
and the $\lambda'_{ijk}$ and $\lambda{''}_{ijk}$ terms
 can contribute to the $b\to s\gamma$, etc~\cite{Barbier:2004ez}.
We would like to point out that the NMSSM with $R$-parity violation has been studied 
before~\cite{Pandita:1999jd},
and the NMSSM with baryon number conservation is similar to 
the $\mu \nu$SSM~\cite{LopezFogliani:2005yw}.

In this paper, to increase the SM-like Higgs boson mass while keep the
sparticle spectrum light, we will concentrate on the natural
and realistic NMSSM with the following properties:
(1) $\tan\beta$ is not large so that the SM-like Higgs boson mass
can be lifted via the tree-level $\lambda S H_d H_u$ term; (2)
The second lightest CP-even Higgs boson
is the SM-like Higgs particle, and then the SM-like Higgs boson mass
can be lifted via the mass matrix diagonalization from Linear Algebra.
However, in such kind of viable parameter space,
the muon $(g_{\mu} - 2)/2$ is generically small due to not large $\tan\beta$.
Thus, to increase muon $(g_{\mu} - 2)/2$,
we need to introduce $R$-parity violation $\lambda_{ijk}$ and
$\lambda'_{ijk}$ terms in Eq.~(\ref{RPV-1}), which will be studied elsewhere.
Interestingly, we may explain the neutrino masses and mixings simultaneously.



\section{$\chi^2$ Analyses for the Phenomenological Constraints and  Experimental Data}

We will consider the $\chi^2$ analyses for the phenomenological
constraints and experimental
data in all three scenarios. For our numerical calculations,
we use the NMSSMTools version 3.2.0~\cite{Ellwanger:2006rn}.

In the original package, the points are survived if they satisfy several phenomenological and theoretical
constraints. Two standard deviation ($95\%$ C.L. upper) limits are applied for those constraints which have
corresponding experimental measurements. In this paper, these two standard deviation ($95\%$ C.L. upper) limits
are replaced by their central values and the experimental errors, which are used
to construct the global $\chi^2$. There are two advantages for this global fit:
(1) The best-fitted benchmark points with minimal $\chi^2$ value can be found exactly, while
the previous method within two standard deviation limits can only provide a
viable parameter space.
(2) The derivation for the central values in the two standard deviation limits can accumulate
to be a relatively significant drift while the global $\chi^2$ can have explicit statistical
meanings for the 1 or 2 standard deviations from the best-fitted points.

In our analyses, several phenomenological and theoretical constraints are
 considered.
These constraints can be divided into the following categories:
\begin{enumerate}
\item  The theoretical constraints and phenomenological constraints,
 which only have $95\%$ C.L. upper limits, are unchanged in the
NMSSMTools~\footnote{For interested readers, the
detailed information can be found in NMSSMTOOLS $3.2.0$, which corresponds to
PROB(1)$\sim$PROB(29), PROB(35), PROB(41)$\sim$ PROB(45), PROB(51) and PROB(52).}.

\item The following LHC Higgs constraints are added:
the second CP-even neutral Higgs field $H_2$ are taken
as the SM-like Higgs boson discovered at the LHC and
its mass is required to be $M_{H_2}\in[124,127]$~GeV.
All the 5 neutral Higgs fields $H_1, H_2, H_3, A_1$ and $A_2$ should satisfy
the LHC constraints, which are taken as the $95\%$ C.L.
on $\sigma/\sigma_{\mbox{\small{SM}}}$
among the LHC measured Higgs mass regions. Table~\ref{LHCupperlimit} shows the ATLAS and CMS Higgs decay
channels that we will consider.

\begin{table}[htb]
\caption{\label{LHCupperlimit} The LHC collider constraints
at $95\%$ C.L. on $\sigma/\sigma_{\mbox{\small{SM}}}$.
7, 8 and 7\&8 means LHC center mass energy $\sqrt{S}=7, ~8$~TeV as well as 7 and 8~TeV combined 
results. The blank means no constraints in this channel. Same conventions are applied in the
 following tables. ``(VH)'' in the table indicates the experimental results are actually measured 
in the vector boson associated production, which is invariant for the WH production channel in the NMSSM.}
\begin{tabular}{ccc}
\hline\hline
 Channels & ATLAS & CMS \\
\hline
$H\to \tau\tau$  &  & 7, 7\&8 \\
$WH\to b b $ & & 7, 7\&8(VH) \\
$H\to b b $ &  & \\
$H\to Z Z$  & 7, 8, 7\&8 & 7, 7\&8\\
$H\to W^+ W^-$ & &7, 7\&8 \\
$H\to \gamma \gamma$  & 7, 8, 7\&8 & 7, 7\&8\\
$2j H\to 2j \gamma\gamma$  & &  \\
\hline
\hline
\end{tabular}
\end{table}

\item  The NMSSMTools two standard deviation constraints
 are replaced by global $\chi^2$ fits, which include:
$b \to s \gamma$, $\delta m_s$, $\delta m_d$, $b \to \tau \nu_\tau$,
$(g_\mu-2)/2$ and Br $(B \to X_s \mu^+\mu^-)$. We update the anomalous
magnetic moment of
the muon $(g_{\mu} - 2)/2$: $\Delta a_{\mu} = a^{\rm exp}_{\mu} - a^{\rm SM}_{\mu}
=(28.7\pm 8.0) \times 10^{-10}$~\cite{Hagiwara:2011af}.

\item LHC Higgs signal strength is constructed in the $\chi^2$, 
as shown in Table~\ref{LHCsignalstrength} and Fig.~\ref{signalstrength-exp}.
Theoretical predictions in the $\chi^2$ correspond to  $H_2$ in the NMSSM.
For the 7 and 8 TeV combined results, the theoretical predicted signal strength for inclusive Higgs production channels are combined from the 7 and 8 TeV individual signal strength proportional to their accumulated luminosities.

\begin{table}[htb]
\caption{\label{LHCsignalstrength}LHC Higgs signal strength in $\chi^2$.}
\begin{tabular}{ccc}
\hline\hline
 Channels & ATLAS & CMS \\
\hline
$H\to \tau\tau$  & 7 & 7\&8 \\
$WH\to b b $     & 7 & 7\&8(VH) \\
$H\to b b $      & 7 & 7\&8\\
$H\to Z Z$       & 7\&8 & 7\&8\\
$H\to W^+ W^-$   & 7 & 7\&8 \\
$H\to \gamma \gamma$  & 7\&8 & 7\&8\\
$2j H\to 2j \gamma\gamma$  & 7\&8 & 7\&8 \\
\hline
\hline
\end{tabular}
\end{table}

\begin{figure}[htbp]
\begin{center}
\includegraphics[width=0.5\textwidth]
{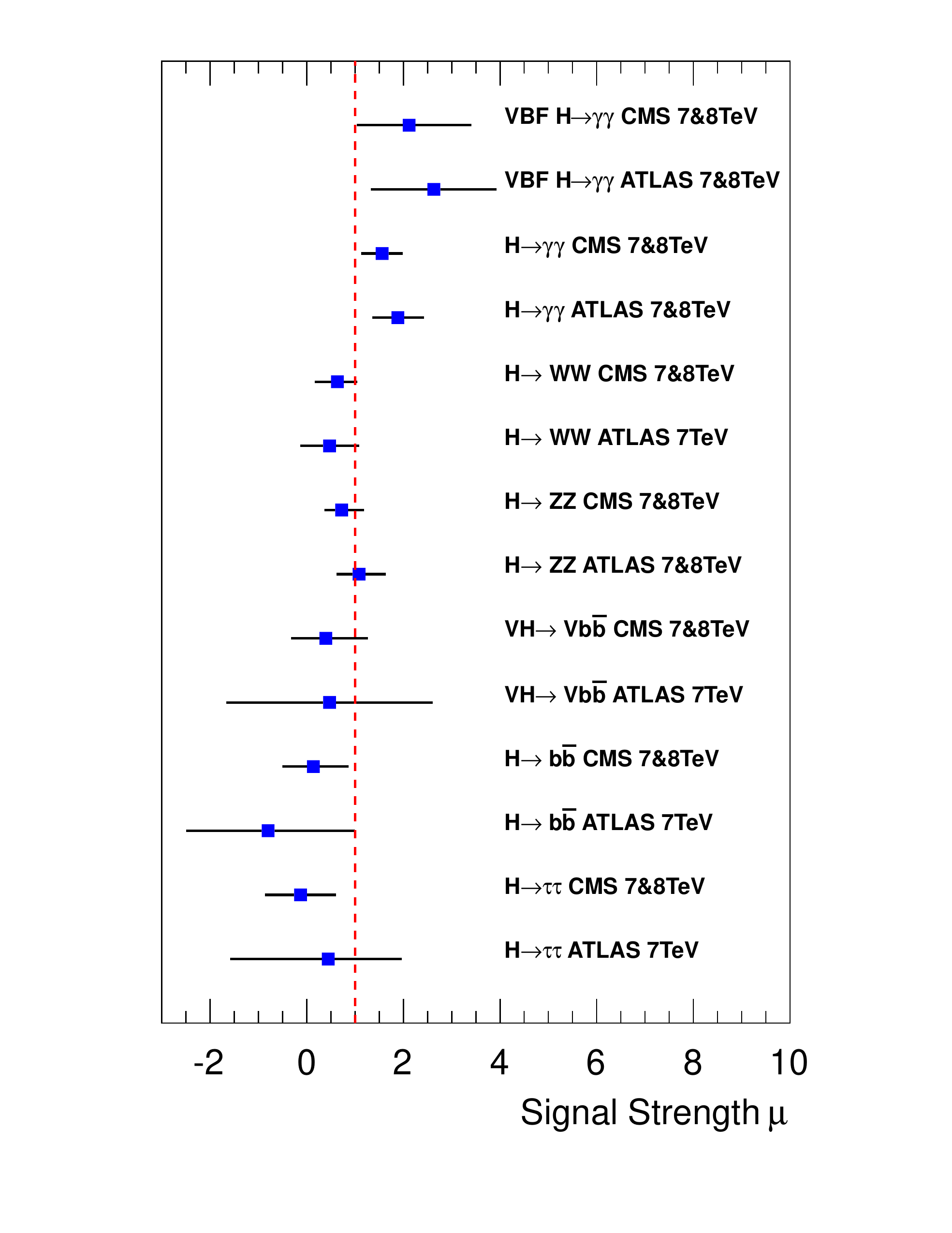}
\end{center}
\caption{\label{signalstrength-exp}LHC Higgs signal strength in different production and decay modes. }
\end{figure}

\item
The cold dark matter relic density is $0.112\pm 0.0056$ from the seven-year
WMAP measurements~\cite{Larson:2010gs}.
For the dark matter, we treat it in three different scenarios, as shown in the Table~\ref{relicdensity}.
In Scenario I, the lightest stable neutralino is required to have the correct dark matter relic density. This is considered in the global $\chi^2$. In Scenario II, the relic density is required to be
smaller than the $95\%$ C.L. experimental upper limit, which assumes
multi-component dark matter. In Scenario III, the relic density is set to be free, which corresponds
to the $R$-parity violation case. Constraints of effective Higgs self-couplings in MicrOMEGAs
and spin independent dark matter direct detection of XENON100 experiment~\cite{Aprile:2011hi} 
in the NMSSMTools package are adjusted according to
different relic density treatment as shown in Table~\ref{relicdensity}.

\begin{table}[htb]
\caption{\label{relicdensity}Three different dark matter relic density
scenarios.
}
\begin{tabular}{lccc}
\hline\hline
& scenario I & scenario II & scenario III\\
\hline
$\Omega \hbar^2$ & in $\chi^2$ & $<95\% \mbox{upper limit}$ & $\mbox{free}$ \\
$\mbox{Effective Higgs Self-Coupling in MicrOMEGAs}>1$ & \checkmark & \checkmark &
$\times$ \\
$\mbox{XENON100}$ & \checkmark & \checkmark & $\times$\\
\hline
\hline
\end{tabular}
\end{table}
\end{enumerate}

To be clear, we have considered three scenarios according to the different dark matter
treatments as explained in category 5.  Categories 1 to 4 are common constraints 
that are applied to all three Scenarios.

The $\chi^2$ is constructed as:
\begin{equation}
\chi^2=\sum\limits_i (\frac{\mu_i^{the}-\mu_i^{exp}}{\delta_i})^2,
\label{equation-chi2}
\end{equation}
in which $\mu_i^{the}$ are theoretical predicted values and
$\mu_i^{exp}$ are corresponding experimental measurements.
$\delta_i$ are one standard fluctuations which includes both statistical
and systematical errors and are taken as the average values for asymmetric errors.

By adopting the above $\chi^2$ constructions, the number of independent variables
in the $\chi^2$ are: 7 in category 3~(Br $(B \to X_s \mu^+\mu^-)$ are considered
 in both low and high dilepton energy regions),  14 LHC Higgs decay signal strength 
(we assume that ATLAS and CMS measurements on the same Higgs decay channels are
 independent), and the different dark matter relic density
Scenarios in category 5. Besides, there are 9 NMSSM input parameters as shown
in Eq.~(\ref{9parameters}). So the number of degree of freedom $n_d$ is 22-9=13 for 
Scenario I and 12 for Scenarios II and III. The goodness of fit can be shown by 
comparing the minimum $\chi^2$ with the $n_d$.

Note that the current top quark mass $m_t$ is $173.5\pm 1$~GeV,
we shall choose the central value $m_t=173.5$ GeV in numerical calculations.
We emphasize that the SM-like Higgs boson $H_2$ mass will increase
and decrease about 1~GeV if we choose the upper limit $m_t=174.5$~GeV
and low limit $m_t=172.5$~GeV, respectively. Thus,
the SM-like Higgs boson $H_2$ mass range from 124~GeV or 127~GeV is fine. Moreover, we define
\begin{eqnarray}
R^{X\overline{X}}_{i}&\equiv& {\sigma(pp\to H_i) \ {\rm BR}(H_i\to X\overline{X})\over
\sigma(pp\to h_{\rm SM)}\ {\rm BR}(h_{\rm SM}\to X\overline{X})}~,~
\end{eqnarray}
where $X\overline{X}$ can be $\gamma \gamma$, $Z^0Z^0$, $W^{+}W^-$,
$b{\bar b}$, and $\tau {\bar \tau}$.





\begin{figure}[htbp]
\begin{center}
\includegraphics[width=0.9\textwidth]
{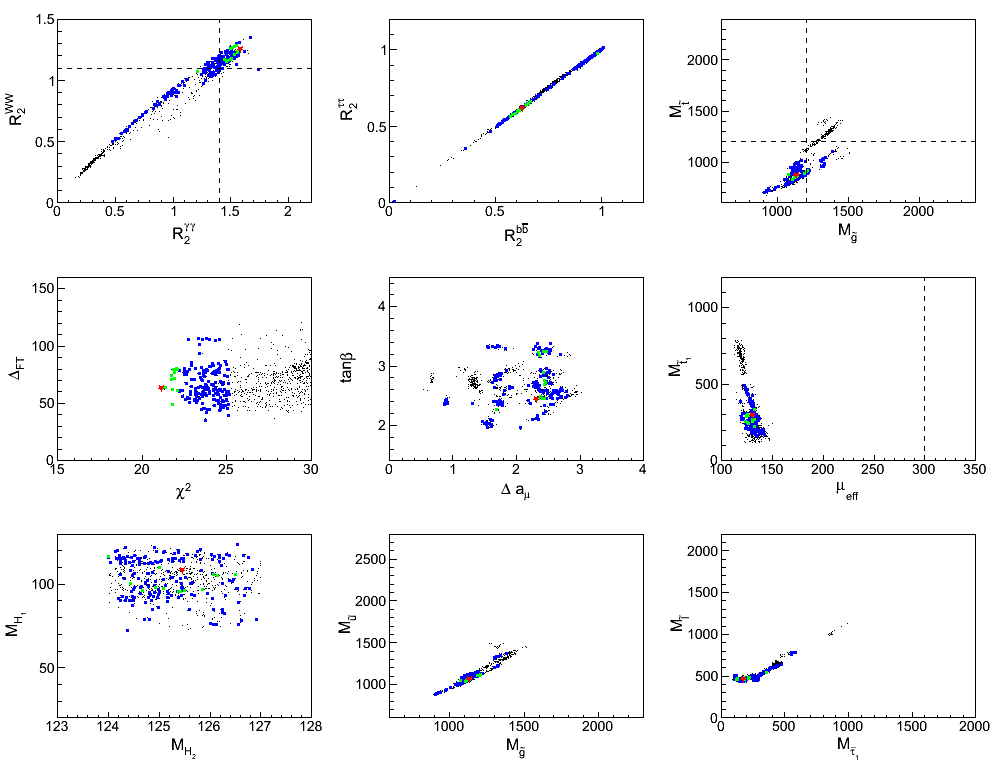}
\end{center}
\caption{\label{Plot9_chi2} The fitting results for Scenario I with relic density 
included in the $\chi^2$. The red stars show the best-fitted benchmark point 
with minimal $\chi_{min}^2=21.16$. The green, blue, and black regions are 
 respectively one, two, and three standard deviation regions 
with $\chi^2<\chi^2_{min}+1, \chi^2_{min}+4$ and $\chi^2_{min}+9$.  }
\end{figure}

\begin{figure}[htbp]
\begin{center}
\includegraphics[width=0.9\textwidth]
{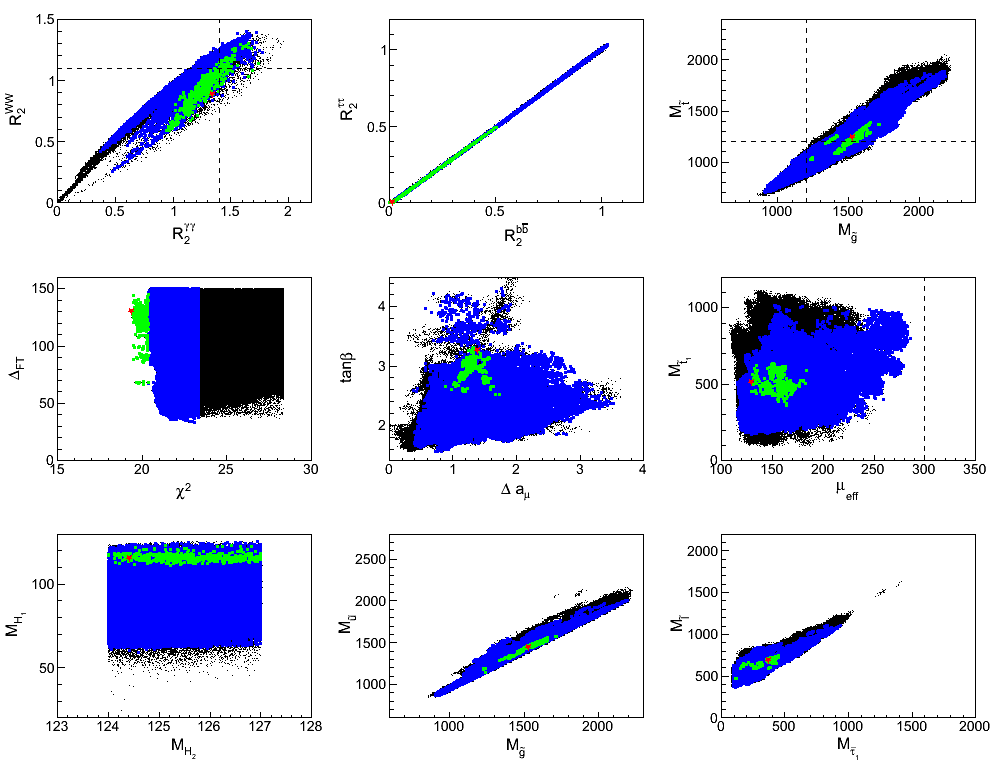}
\end{center}
\caption{\label{Plot9_uplimit} The fitting results for Scenario II with relic density 
smaller than the $95\%$ C.L. upper limit. The red stars show the best-fitted benchmark 
point with minimal $\chi_{min}^2=19.35$. The green, blue, and black regions are respectively 
one, two, and three standard deviation regions 
with $\chi^2<\chi^2_{min}+1, ~\chi^2_{min}+4$, and $\chi^2_{min}+9$.  }
\end{figure}

\begin{figure}[htbp]
\begin{center}
\includegraphics[width=0.9\textwidth]
{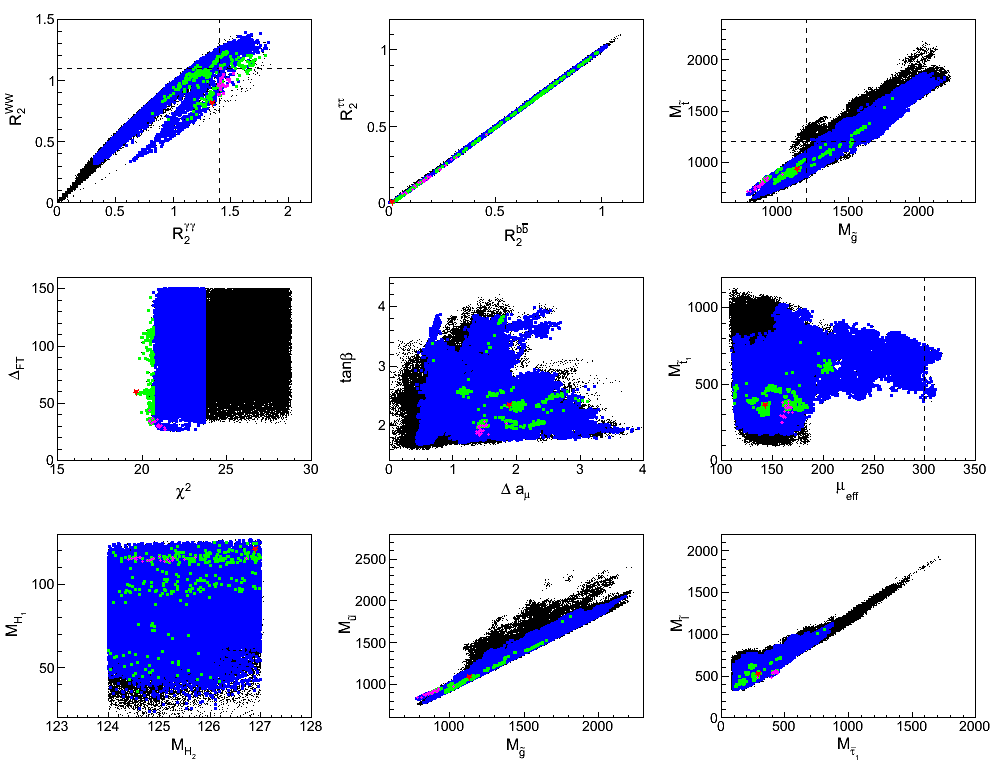}
\end{center}
\caption{\label{Plot9_free} The fitting results for Scenario III without $R$-parity. 
The red stars show the best-fitted benchmark point with minimal $\chi_{min}^2=19.67$. 
The green, blue, and black regions are respectively one, two, and three standard 
deviation regions with $\chi^2<\chi^2_{min}+1,~ \chi^2_{min}+4$, and $\chi^2_{min}+9$. 
The magenta region corresponds to $R_{\gamma\gamma}>1.4$, $R_{VV}<1.1$, $R_{bb}<1.0$, 
$R_{\tau\tau}<1.0$, $M_{\tilde t}=\sqrt{m_{\widetilde{t}_1}^2+m_{\widetilde{t}_2}^2}<1.2{\rm TeV}$,
 $\mu_{\rm eff}<300$~GeV, $M_{\widetilde{g}}<1.2$~TeV, $\chi^2<\chi^2_{min}+4$,
 and $\Delta_{\rm FT}<50$. }
\end{figure}

 We present the $R_2^{VV}$ versus $R_2^{\gamma\gamma}$,
$R_2^{\tau\bar{\tau}}$ versus $R_2^{b{\bar b}}$,
$M_{\tilde t}$ versus $M_{\tilde g}$, $\Delta_{\rm FT}$
versus $\chi^2$, $\tan\beta$ versus $\Delta a_{\mu}$,
$m_{{\tilde t}_1} $ versus $\mu_{\rm eff}$,
 $M_{H_1}$ versus $M_{H_2}$, 
$M_{\tilde u}$ versus $M_{\tilde g}$,
and $M_{\tilde l}$ versus $M_{{\tilde \tau}_1}$  in
Figs. \ref{Plot9_chi2}, \ref{Plot9_uplimit}, and \ref{Plot9_free}
for Scenarios I, II, and III, respectively.
 The red stars show the best-fitted benchmark points 
with minimal $\chi_{min}^2=21.16, ~19.35, ~19.67$
 for Scenarios I, II, III, respectively.  The magenta region corresponds to
 $R_{\gamma\gamma}>1.4$, $R_{VV}<1.1$, $R_{bb}<1.0$, $R_{\tau\tau}<1.0$,
 $M_{\tilde t}=\sqrt{m_{\widetilde{t}_1}^2+m_{\widetilde{t}_2}^2}<1.2~{\rm TeV}$,
 $\mu_{\rm eff}<300$~GeV, $M_{\widetilde{g}}<1.2$~TeV, $\chi^2<\chi^2_{min}+4$,
 and $\Delta_{\rm FT}<50$. In particular, the small $\chi^2$ and $\Delta_{\rm FT} < 50$
 are not compatible with each other in Scenarios I and II due to 
XENON100 experimental constraint, and then only Scenario III has magenta region.
In addition to the  minimal $\chi^2$ points, we consider three kinds of other 
benchmark points:
\begin{itemize}

\item Benchmark points IA, IIA, and IIIA have small $\chi^2$, relatively small $\Delta_{\rm FT}$,
the light stop mass around 200~GeV, and the first two generation squark masses lighter
than about 1.1~TeV.

\item Benchmark points IB, IIB, and IIIB have  small $\chi^2$, relatively small $\Delta_{\rm FT}$,
the light stop mass around 200~GeV, and the first two generation squark masses heavier
than about 1.1~TeV. 

\item Benchmark points IC, IIC, and IIIC have  small $\chi^2$, relatively small $\Delta_{\rm FT}$,
and relatively heavier light stop.

\end{itemize}
We present the minimal $\chi^2$ point, and three other benchmark points
in Tables~\ref{benchmarkpttable-I}, \ref{benchmarkpttable-II}, and
\ref{benchmarkpttable-III} for Scenarios I, II, and III, respectively.
Moreover, we study the constraints from the LHC supersymmetry searches,
and find that only the benchmark points IA and IIA are excluded
by the current LHC supersymmetry searches~\cite{Aad:2011ib, Aad:2011qa, 
Chatrchyan:2011zy, SUSY-ATLAS-CMS, LHC-SUSY-EX}. Because the benchmark points
I$\chi^2_{min}$, IC, IIB and IIC have relatively light spectra 
(e.g.~less thank about 1.2~TeV),
 the LHC supersymmetry search constraints are relaxed in our models,
which will be explained briefly in the following.
The benchmark points in Scenarios I and II have fine-tuning from
about 1\% to  2\%, while the benchmark points in Scenario III have
fine-tuning from about 2\% to 3.7\%.
In particular, in the best benchmark point IIIA we have $\chi^2=21.31$, 
and $\Delta_{\rm FT}=27.0$, {\it i.e.}, 3.7\% fine-tuning.
Also, all the supersymmetric particles are lighter than 830~GeV.
By the way, all the benchmark points except II$\chi^2_{min}$ satisfy
the naturalnes conditions: $\mu_{\rm eff} < 300$~GeV, $M_{\tilde t}<1.2$~TeV,
and $M_{\tilde g} \le 1.5$ TeV. Note that $\Delta_{FT}=130.4$ in benchmark
point II$\chi^2_{min}$, the two 
fine-tuning definitions in Section II are compatible.

From  the viable parameter space in
Figs. \ref{Plot9_chi2}, \ref{Plot9_uplimit}, and \ref{Plot9_free}, we find that
$\tan\beta$ is generically smaller than about 4.5, and then 
we have the small anomalous magnetic moment of the muon $(g_{\mu} - 2)/2$,
{\it i.e.}, $\Delta a_{\mu} < 4.0\times 10^{-10}$. 
Also, we notice
the correlation between $R_2^{\gamma\gamma}$ and $R_2^{VV}$ which
roughly is $R_2^{\gamma\gamma} \sim 1.27 \times R_2^{VV}$.
Interestingly, we do have some viable parameter space which indeed
have $R_2^{\gamma\gamma} \ge 1.4$ and $R_2^{VV} \le 1.1$.
The generic features for the parameter space with small $\chi^2$ are that
the light stop is around 500 GeV or smaller,  the singlino and Higgsino are light
 neutralinos and chargino, the Wino-like chargino is heavy, and
the Bino-like and Wino-like neutralinos are
the second heaviest neutralino and the heaviest neutralino, respectively.
Thus, we can understand why the LHC supersymmetry search constraints
are relaxed: the branch ratios of the first two generation
squarks decaying directly to the LSP neutralino and quarks are very small
around 1\%, and the dominant decay channels are Wino-like
chargino/neutralino and quarks. And then the
Wino-like chargino and neutralino will decay into quite a few jets
or leptons via the light chargino and neutralinos. Also, gluino will
decay dominant into the stop and top
 quarks, which have long decay chains as well.
 Therefore, the LHC supersymmetry search
constraints can be  relaxed. The detailed LHC supersymmetry
search constraints will be studied elsewhere.
Moreover, because the LSP neutralino has relatively large Higgsino
components due to small effective $\mu$ term, the XENON100 experiment
gives strong constraint, for example, the spin-independent 
LSP neutralino-nucleon cross sections are larger than the
XENON100 experiment upper bound in the bencharmark points III$\chi^2_{min}$, IIIA,
and IIIC~\cite{Aprile:2011hi} if $R$-parity is conserved. 
This is another reason why we get better points in
Scenario III than Scenario II.
In Scenario III, the constraints from the LHC supersymmetry searhes
and XENON100 experiment can be escaped, and  the
$R$-parity violating $ \lambda_{ijk}$ and $\lambda'_{ijk}$ terms can
increase $(g_{\mu} - 2)/2$ and generate the neutrino masses and mixings.
Therefore, Scenario III with  $R$-parity violation is more natural and
realistic than Scenarios I and II.


\begin{table}[ht]
\begin{center}

\begin{tabular}{rl}

\begin{tabular}{|c|c|c|c|c|} \hline
Point &   I$\chi^2_{min}$ & IA & IB & IC \\
\hline 
$M_0$ & 	264& 	294& 	562& 	249    \\
$M_{1/2}$ & 	489& 	484& 	624& 	571    \\
$\tan \beta$ &	2.436& 	2.851& 	2.910& 	3.293    \\
$\lambda$&	0.601& 	0.550& 	0.564& 	0.536    \\
$\kappa$  &	0.245& 	0.249& 	0.268& 	0.243    \\
$A_0$ &	-1180& 	-1253& 	-1779& 	-1441    \\
$A_\lambda$ &	-315& 	-230& 	-628& 	-193    \\
$A_\kappa$ &	-1.904& 	-2.028& 	-218.408& 	-1.900    \\
$\mu_{\mbox{eff}}$ &	130& 	124& 	123& 	128    \\
$M_1$ & 207& 204& 266& 242 \\
$M_2$ & 384& 380& 493& 449\\
$M_3$ & 1091& 1081& 1368& 1259\\
$\widetilde{\chi}_{1}^{0}$& 	75& 	70& 	77& 	76    \\
$\widetilde{\chi}_{2}^{0}$& 	163& 	-156& 	-156& 	-158    \\
$\widetilde{\chi}_{3}^{0}$& 	-168& 	163& 	172& 	171    \\
$\widetilde{\chi}_{4}^{0}$& 	219& 	216& 	273& 	250    \\
$\widetilde{\chi}_{5}^{0}$& 	415& 	410& 	521& 	475    \\
$\widetilde{\chi}_{1}^{\pm}$&	114& 	109& 	113& 	117    \\
$\widetilde{\chi}_{2}^{\pm}$&	414& 	409& 	520& 	475    \\
$\widetilde{g}$ &	1134& 	1125& 	1436& 	1305    \\
$\widetilde{\nu}_{e/\mu}$&	457& 	472& 	741& 	497    \\
$\widetilde{\nu}_{\tau}$&	456& 	471& 	740& 	496    \\
$\widetilde{e}_{R}/\widetilde{\mu}_{R}$&	178& 	216& 	482& 	126    \\
$\widetilde{e}_{L}/\widetilde{\mu}_{L}$&	461& 	477& 	744& 	502    \\
$\widetilde{\tau}_{1}$&	176& 	213& 	479& 	116    \\
$\widetilde{\tau}_{2}$&	461& 	476& 	744& 	501    \\
$\chi^2$ & 	21.16& 	24.85& 	25.92& 	23.37    \\
\hline
\end{tabular}
&
\begin{tabular}{|c|c|c|c|c|} \hline
Point &   I$\chi^2_{min}$ &IA & IB & IC \\
\hline
$\widetilde{t}_{1}$&	294& 	210& 	186& 	257    \\
$\widetilde{t}_{2}$&	822& 	810& 	1026& 	920    \\
$\widetilde{t}$&	873& 	837& 	1042& 	955    \\
$\widetilde{b}_{1}$&	783& 	768& 	994& 	880    \\
$\widetilde{b}_{2}$&	983& 	979& 	1296& 	1114    \\
$\widetilde{u}_{R}/\widetilde{c}_{R}$&	1053& 	1052& 	1397& 	1197    \\
$\widetilde{u}_{L}/\widetilde{c}_{L}$&	1059& 	1057& 	1398& 	1206    \\
$\widetilde{d}_{R}/\widetilde{s}_{R}$&	1014& 	1013& 	1342& 	1153    \\
$\widetilde{d}_{L}/\widetilde{s}_{L}$&	1061& 	1059& 	1399& 	1208    \\
$H_1^0$&	108.5& 	95.5& 	111.7& 	91.7    \\
$H_2^0$&	125.5& 	125.4& 	125.2& 	124.2    \\
$H_3^0$&	359.7& 	396.3& 	385.6& 	461.1    \\
$A_1$&	99.6& 	123.8& 	91.7& 	138.4    \\
$A_2$&	353.4& 	390.0& 	377.9& 	455.9    \\
$H^\pm$&	343.9& 	383.2& 	371.3& 	450.2    \\
$\Omega \hbar^2$ &	0.110& 	0.104& 	0.103& 	0.109    \\
$\Delta_{a_\mu}$ [$10^{-10}$]&	2.317& 	2.586& 	1.157& 	1.823    \\
$\sigma^{si}(p)$ [$10^{-10}$~pb]&	7.468& 	1.137& 	3.208& 	39.683    \\
${\rm Br}^{(b \rightarrow s\gamma)}$ [$10^{-4}$] &	3.342& 	2.633& 	2.714& 	2.747    \\
$\Delta_{\rm FT}$ &	62.7& 	74.0& 	109.3& 	101.5    \\
${R_2}_{\rm VBF}^{\gamma\gamma}$ &	1.48& 	1.60& 	1.75& 	1.46    \\
${R_2}^{\gamma\gamma}$ &	1.58& 	1.43& 	1.45& 	1.31    \\
${R_2}^{WW}$ &	1.25& 	1.10& 	1.07& 	1.09    \\
${R_2}^{ZZ}$ &	1.25& 	1.10& 	1.07& 	1.09    \\
${R_2}^{Vbb}$ &	0.59& 	0.59& 	0.50& 	0.70    \\
${R_2}^{bb}$ &	0.63& 	0.53& 	0.41& 	0.63    \\
${R_2}^{\tau\tau}$ &	0.62& 	0.53& 	0.41& 	0.62    \\
\hline
\end{tabular}
\\
\end{tabular}
\end{center}
\caption{Particle spectra (in GeV) and parameters for benchmark points in Scenario I.}		
\label{benchmarkpttable-I}
\end{table}

\begin{table}[ht]
\begin{center}

\begin{tabular}{rl}

\begin{tabular}{|c|c|c|c|c|} \hline
Point &   II$\chi^2_{min}$ & IIA & IIB & IIC \\
\hline
$M_0$ & 	454& 	208& 	576& 	279    \\
$M_{1/2}$ & 	673& 	413& 	528& 	527    \\
$\tan \beta$ &	3.267& 	2.518& 	2.468& 	1.820    \\
$\lambda$&	0.452& 	0.617& 	0.582& 	0.584    \\
$\kappa$  &	0.214& 	0.295& 	0.247& 	0.172    \\
$A_0$ &	-1520& 	-1090& 	-1519& 	-986    \\
$A_\lambda$ &	-296& 	-248& 	-618& 	-380    \\
$A_\kappa$ &	-1.354& 	-1.049& 	-253.408& 	-1.707    \\
$\mu_{\mbox{eff}}$ &	130& 	129& 	118& 	123    \\
$M_1$ & 287& 173&224& 224\\
$M_2$ & 530& 324&416& 416\\
$M_3$ & 1467& 933 & 1169& 1173\\
$\widetilde{\chi}_{1}^{0}$& 	87& 	70& 	69& 	74    \\
$\widetilde{\chi}_{2}^{0}$& 	-152& 	-166& 	-156& 	133    \\
$\widetilde{\chi}_{3}^{0}$& 	167& 	170& 	158& 	-166    \\
$\widetilde{\chi}_{4}^{0}$& 	292& 	194& 	233& 	233    \\
$\widetilde{\chi}_{5}^{0}$& 	557& 	358& 	446& 	444    \\
$\widetilde{\chi}_{1}^{\pm}$&	122& 	107& 	104& 	107    \\
$\widetilde{\chi}_{2}^{\pm}$&	556& 	357& 	446& 	444    \\
$\widetilde{g}$ &	1529& 	968& 	1235& 	1218    \\
$\widetilde{\nu}_{e/\mu}$&	680& 	377& 	710& 	489    \\
$\widetilde{\nu}_{\tau}$&	679& 	377& 	709& 	489    \\
$\widetilde{e}_{R}/\widetilde{\mu}_{R}$&	376& 	122& 	510& 	191    \\
$\widetilde{e}_{L}/\widetilde{\mu}_{L}$&	683& 	383& 	713& 	492    \\
$\widetilde{\tau}_{1}$&	372& 	118& 	509& 	190    \\
$\widetilde{\tau}_{2}$&	683& 	383& 	712& 	492    \\
$\chi^2$ & 	19.35& 	24.19& 	23.86& 	23.70    \\
\hline
\end{tabular}

\begin{tabular}{|c|c|c|c|c|} \hline
Point &   II$\chi^2_{min}$ &IIA & IIB & IIC \\
\hline
$\widetilde{t}_{1}$&	513& 	157& 	184& 	432    \\
$\widetilde{t}_{2}$&	1126& 	700& 	912& 	895    \\
$\widetilde{t}$&	1237& 	717& 	930& 	994    \\
$\widetilde{b}_{1}$&	1093& 	654& 	880& 	866    \\
$\widetilde{b}_{2}$&	1348& 	835& 	1153& 	1058    \\
$\widetilde{u}_{R}/\widetilde{c}_{R}$&	1434& 	898& 	1240& 	1127    \\
$\widetilde{u}_{L}/\widetilde{c}_{L}$&	1445& 	903& 	1237& 	1135    \\
$\widetilde{d}_{R}/\widetilde{s}_{R}$&	1384& 	865& 	1192& 	1086    \\
$\widetilde{d}_{L}/\widetilde{s}_{L}$&	1446& 	905& 	1239& 	1136    \\
$H_1^0$&	115.2& 	106.5& 	99.7& 	98.0    \\
$H_2^0$&	124.4& 	126.2& 	124.7& 	126.0    \\
$H_3^0$&	440.5& 	363.1& 	332.7& 	290.1    \\
$A_1$&	66.1& 	145.4& 	103.3& 	74.6    \\
$A_2$&	435.0& 	355.4& 	325.3& 	288.4    \\
$H^\pm$&	433.6& 	345.6& 	316.3& 	275.7    \\
$\Omega \hbar^2$ &	0.038& 	0.001& 	0.072& 	0.070    \\
$\Delta_{a_\mu}$ [$10^{-10}$]&	1.385& 	3.410& 	1.100& 	1.513    \\
$\sigma^{si}(p)$ [$10^{-10}$~pb]&	59.671& 	39.947& 	22.116& 	28.064    \\
${\rm Br}^{(b \rightarrow s\gamma)}$ [$10^{-4}$] &	3.553& 	2.294& 	2.914& 	4.123    \\
$\Delta_{\rm FT}$ &	130.4& 	48.8& 	75.7& 	59.6    \\
${R_2}_{\rm VBF}^{\gamma\gamma}$ &	1.08& 	1.68& 	1.53& 	1.24    \\
${R_2}^{\gamma\gamma}$ &	1.34& 	1.45& 	1.42& 	1.42    \\
${R_2}^{WW}$ &	0.89& 	1.09& 	1.10& 	1.09    \\
${R_2}^{ZZ}$ &	0.89& 	1.09& 	1.10& 	1.09    \\
${R_2}^{Vbb}$ &	0.01& 	0.57& 	0.72& 	0.65    \\
${R_2}^{bb}$ &	0.01& 	0.49& 	0.67& 	0.74    \\
${R_2}^{\tau\tau}$ &	0.00& 	0.49& 	0.66& 	0.73    \\

\hline
\end{tabular}
\\
\end{tabular}
\end{center}
\caption{Particle spectra (in GeV) and parameters  for benchmark points in Scenario II.}		
\label{benchmarkpttable-II}
\end{table}

\begin{table}[ht]
\begin{center}

\begin{tabular}{rl}

\begin{tabular}{|c|c|c|c|c|} \hline
Point &   III$\chi^2_{min}$ & IIIA & IIIB & IIIC \\
\hline
$M_0$ & 	352& 	431& 	344& 	337    \\
$M_{1/2}$ & 	489& 	326& 	500& 	441    \\
$\tan \beta$ &	2.341& 	1.853& 	2.731& 	2.039    \\
$\lambda$&	0.604& 	0.589& 	0.623& 	0.614    \\
$\kappa$  &	0.295& 	0.364& 	0.288& 	0.318    \\
$A_0$ &	-1063& 	-372& 	-1322& 	-620    \\
$A_\lambda$ &	-329& 	-161& 	-267& 	-9.93$\times10^{-6}$    \\
$A_\kappa$ &	-128.134& 	-628.214& 	-0.877& 	-1.459    \\
$\mu_{\mbox{eff}}$ &	144& 	162& 	150& 	166    \\
$M_1$ & 207& 136& 212& 186\\
$M_2$ & 384& 255& 393& 346\\
$M_3$ & 1090& 745& 1114& 990\\
$\widetilde{\chi}_{1}^{0}$& 	91& 	86& 	92& 	107    \\
$\widetilde{\chi}_{2}^{0}$& 	-177& 	160& 	-184& 	195    \\
$\widetilde{\chi}_{3}^{0}$& 	187& 	-189& 	191& 	-197    \\
$\widetilde{\chi}_{4}^{0}$& 	222& 	232& 	228& 	221    \\
$\widetilde{\chi}_{5}^{0}$& 	417& 	310& 	425& 	384    \\
$\widetilde{\chi}_{1}^{\pm}$&	126& 	122& 	133& 	141    \\
$\widetilde{\chi}_{2}^{\pm}$&	416& 	306& 	425& 	383    \\
$\widetilde{g}$ &	1138& 	796& 	1162& 	1035    \\
$\widetilde{\nu}_{e/\mu}$&	513& 	502& 	514& 	475    \\
$\widetilde{\nu}_{\tau}$&	512& 	502& 	514& 	475    \\
$\widetilde{e}_{R}/\widetilde{\mu}_{R}$&	291& 	395& 	273& 	289    \\
$\widetilde{e}_{L}/\widetilde{\mu}_{L}$&	517& 	505& 	519& 	479    \\
$\widetilde{\tau}_{1}$&	289& 	395& 	270& 	288    \\
$\widetilde{\tau}_{2}$&	517& 	505& 	518& 	479    \\
$\chi^2$ & 	19.67& 	21.31& 	23.85& 	20.53    \\
\hline
\end{tabular}

\begin{tabular}{|c|c|c|c|c|} \hline
Point &   III$\chi^2_{min}$ &IIIA & IIIB & IIIC \\
\hline
$\widetilde{t}_{1}$&	365& 	252& 	184& 	391    \\
$\widetilde{t}_{2}$&	850& 	647& 	832& 	795    \\
$\widetilde{t}$&	925& 	694& 	852& 	886    \\
$\widetilde{b}_{1}$&	814& 	610& 	793& 	760    \\
$\widetilde{b}_{2}$&	1012& 	778& 	1020& 	929    \\
$\widetilde{u}_{R}/\widetilde{c}_{R}$&	1078& 	827& 	1097& 	985    \\
$\widetilde{u}_{L}/\widetilde{c}_{L}$&	1083& 	823& 	1101& 	990    \\
$\widetilde{d}_{R}/\widetilde{s}_{R}$&	1039& 	798& 	1056& 	952    \\
$\widetilde{d}_{L}/\widetilde{s}_{L}$&	1085& 	825& 	1103& 	992    \\
$H_1^0$&	120.5& 	114.3& 	119.6& 	116.6    \\
$H_2^0$&	126.9& 	124.6& 	124.9& 	125.7    \\
$H_3^0$&	367.8& 	342.7& 	436.6& 	382.5    \\
$A_1$&	154.3& 	300.6& 	157.1& 	247.6    \\
$A_2$&	360.0& 	335.1& 	430.1& 	376.3    \\
$H^\pm$&	352.6& 	330.2& 	421.7& 	369.3    \\
$\Omega \hbar^2$ &$\times$& 	$\times$& 	$\times$& 	$\times$    \\
$\Delta_{a_\mu}$ [$10^{-10}$]&	1.893& 	1.587& 	2.207& 	1.848    \\
$\sigma^{si}(p)$ [$10^{-10}$~pb]&	$\times$& 	$\times$& 	$\times$& 	$\times$    \\
${\rm Br}^{(b \rightarrow s\gamma)}$ [$10^{-4}$] &	3.586& 	3.413& 	2.560& 	3.659    \\
$\Delta_{\rm FT}$ &	59.5& 	27.0& 	68.4& 	44.1    \\
${R_2}_{\rm VBF}^{\gamma\gamma}$ &	0.88& 	0.93& 	1.60& 	1.16    \\
${R_2}^{\gamma\gamma}$ &	1.34& 	1.46& 	1.41& 	1.42    \\
${R_2}^{WW}$ &	0.81& 	0.95& 	1.08& 	1.10    \\
${R_2}^{ZZ}$ &	0.81& 	0.95& 	1.08& 	1.10    \\
${R_2}^{Vbb}$ &	0.01& 	0.10& 	0.55& 	0.37    \\
${R_2}^{bb}$ &	0.02& 	0.16& 	0.48& 	0.46    \\
${R_2}^{\tau\tau}$ &	0.00& 	0.14& 	0.47& 	0.44    \\
\hline
\end{tabular}
\\
\end{tabular}
\end{center}
\caption{Particle spectra (in GeV) and parameters  for benchmark points in Scenario III.}		
\label{benchmarkpttable-III}
\end{table}




\section{Conclusion}


We pointed out that as a solution to the SM fine-tuning problems,
 supersymmetry needs not to provide the dark matter candidate, {\it i.e.},
$R$-parity can be violated. Because the NMSSM can explain the Higgs boson mass
and decay rates better than the MSSM, we considered three kinds of the NMSSM scenarios.
 To satisfy the phenomenological constraints and fit the experimental
data, we studied the $\chi^2$ analyses for all three kinds of Scenarios, and showed
that the Higgs boson mass and decay rates can indeed be explained very well.
 For the small $\chi^2$ values and fine-tuning around 2-3.7\%,
we obtained the viable parameter space with light (e.g. less than about 900 GeV) 
supersymmetric particle spectra only in Scenario III.
With the small  $\chi^2$ values and fine-tuning around 1-2\%,
we got the viable parameter space
with relatively heavy (e.g. less than about 1.2~TeV) supersymmetric particle spectra.
Especially,
the best benchmark point is IIIA, which has almost minimal $\chi^2$ and 3.7\% fine-tuning.
The correct dark matter density can be realized in Scenario I as well.
The generic features for the viable parameter space with smaller $\chi^2$ are that
the light stop is around 500 GeV or smaller,  the singlino and Higgsino are light
chargino and neutralinos, the Wino-like chargino is heavy, and
the Bino-like and Wino-like neutralinos are
the second heaviest neutralino and heaviest neutralinos, respectively.
Thus, we found that the LHC supersymmetry
search constraints can be relaxed due to quite a few jets and/or leptons in
the final states in Scenarios I and II, but the XENON100 experiment still gives
strong constraint on the dark matter direct detection cross sections. 
Moreover, $\tan\beta$ is not large and the second lightest CP-even
 Higgs particle is SM-like so that the SM-like Higgs boson mass can be lifted.
However, the extra contributions to the muon $(g_{\mu} - 2)/2$
 are smaller than three sigma low bound in general due to
relatively small $\tan\beta$.
With $R$-parity violation, we can evade the
LHC supersymmetry search and XENON100 experiment
constraints, and the $R$-parity violation superpotential
term(s) may increase $(g_{\mu} - 2)/2$ and explain the neutrino masses
and mixings. Therefore,  Scenario III with  $R$-parity violation is more natural and
realistic than Scenarios I and II.

\begin{acknowledgments}

We would like to thank Zhaofeng Kang and Chunli Tong for
collaboration in the early stage of this project.
This research was supported in part
by the Natural Science Foundation of China
under grant numbers 11075003, 10821504, 11075194, and 11135003,
by the Postdoctoral Science Foundation of China under grant
number Y2Y2231B11, 
and by the DOE grant DE-FG03-95-Er-40917.

\end{acknowledgments}


\end{document}